\documentclass[journal=jacsat,manuscript=article]{achemso}

\usepackage[version=3]{mhchem} 

\author{Chang-Chun He}
\affiliation{Department of Physics, Southern University of Science and Technology, Shenzhen 518055, China}
\alsoaffiliation{Department of Physics, South China University of Technology, Guangzhou 510640, China}
\author{Shao-Gang Xu}
\affiliation{Department of Physics, Southern University of Science and Technology, Shenzhen 518055, China}
\author{Shao-Bin Qiu}
\affiliation{Department of Physics, South China University of Technology, Guangzhou 510640, China}
\author{Chao He}
\affiliation{Department of Physics, Southern University of Science and Technology, Shenzhen 518055, China}
\author{Yu-Jun Zhao}
\affiliation{Department of Physics, South China University of Technology, Guangzhou 510640, China}
\author{Xiao-Bao Yang}
\affiliation{Department of Physics, South China University of Technology, Guangzhou 510640, China}
\email{scxbyang@scut.edu.cn}
\author{Hu Xu}
\affiliation{Department of Physics, Southern University of Science and Technology, Shenzhen 518055, China}
\alsoaffiliation{Shenzhen Key Laboratory of Advanced Quantum Functional Materials and Devices and Department of Physics, Southern University of Science and Technology, China}
\email{xuh@sustech.edu.cn}

\title{Five-fold Symmetry in Au-Si  Metallic Glass}

\begin{document}

\begin{tocentry}

\includegraphics[width=8.3cm]{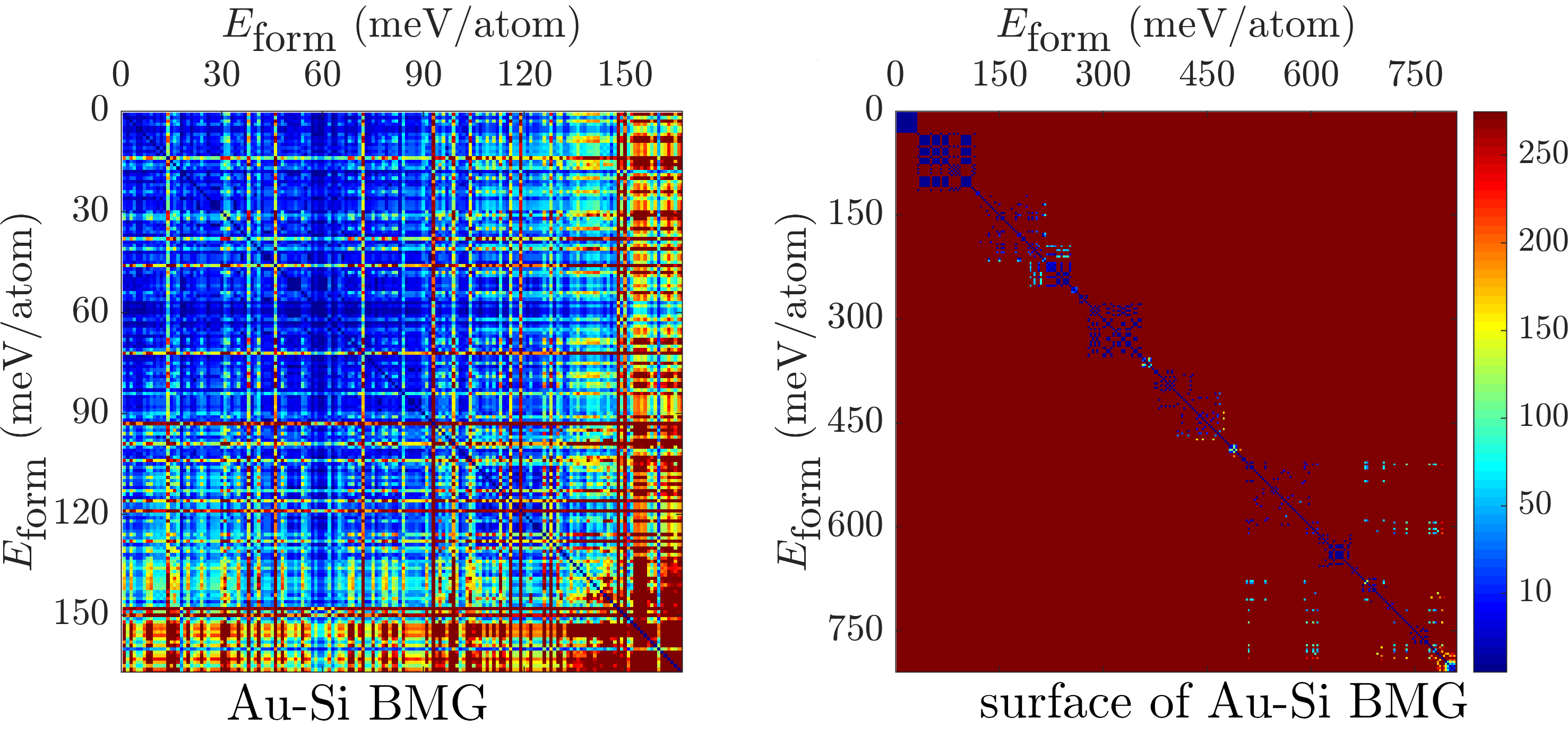}

\end{tocentry}

\begin{abstract}
The first metallic glass of Au-Si alloy has been discovered for over half a century, but its atomic structure is still puzzling. Herein, \ce{Au8Si} dodecahedrons with local five-fold symmetry are revealed as building blocks in Au-Si metallic glass, and the interconnection modes of \ce{Au8Si} dodecahedrons determine  the medium-range order. With dimensionality reduction, the surface ordering is  attributed to the motif transformation of \ce{Au8Si} dodecahedrons into planar \ce{Au5Si} pyramids with five-fold symmetry, and thus the self-assembly of \ce{Au5Si} pyramids leads to the formation of the ordered \ce{Au2Si} monolayer with the lowest energy.  Furthermore, the structural similarity analysis is performed to unveil the physical origin of structural characteristics in different dimensions. The amorphism of Au-Si is due to the smooth energy landscape around the global minimum, while the ordered surface structure occurs due to the steep energy landscape.
\end{abstract}

\section{Introduction}

The Au-Si alloy was the first reported metallic glass, which was discovered in 1960 by rapid cooling \cite{KLEMENT1960}. Subsequently, bulk metallic glasses (BMGs) have attracted increasing attention due to their fundamental scientific interests and practical applications \cite{PhysRevLett.103.075502,INOUE2000279,doi:10.1126/science.1232450}. 
Up to now, many computational methods and experimental techniques have been involved to investigate structures of BMGs \cite{PhysRevLett.90.195504,Hirata2011,Zhong2014}.  However, due to the absence of long-range order it remains challenging to determine atomic structures of BMGs \cite{PhysRevLett.102.245501}. For example, a combination of experimental measurements and reverse Monte Carlo modelings failed to reproduce the precise packing structures of BMGs at the atomic level without sufficient structural information \cite{Keen1990}. Numerous structural models were proposed to improve the understanding of short-range order (SRO) in BMGs over the years \cite{BERNAL1960, GASKELL1978}, but these models are still difficult to figure out the  medium-range order (MRO) of BMGs \cite{Sheng2006}. To overcome this challenge, the efficient cluster packing  on a face-centered cubic lattice was proposed \cite{Miracle2004}. In 2009, the MRO in BMGs was described by the packing mode of local motifs on a network with fractal dimension ($D_f = 2.32$)  \cite{Ma2009}, which is smaller than that of crystal ($D_f = 3.0$) or  dodecahedron-quasicrystal ($D_f = 2.72$), indicating that  filling the real space in these amorphous solids by building blocks is impossible. Therefore, unveiling the nature of atomic packing at the MRO scale in BMGs \cite{Miracle2007_MRS_Bu} is a long-standing problem.

The Au-Si eutectic alloy has the extremely low eutectic temperature (359 $^{\circ}$C) below the melting points of Au (1063 $^{\circ}$C) and Si (1412 $^{\circ}$C) \cite{Okamoto1983}.  At the lowest eutectic temperature, it has a composition of \ce{Au82Si18} \cite{Okamoto1983}, which  is the key to the catalytic growth of silicon nanowires \cite{Hannon2006} because Si atoms maintain relatively high mobility at low temperatures \cite{PhysRevB.81.140202}. In recent years, although some theoretical studies reported possible alloy structures of Au-Si \cite{TASCI2010449,PhysRevB.95.134109,doi:10.1063/1.2815326}, the local ordering of Au-Si BMG is still lacking. Interestingly, two-dimensional (2D) crystallization was observed at the  surface of Au-Si metallic glass \cite{Shpyrko77,https://doi.org/10.1002/advs.201903544}, which has been experimentally confirmed to be an ordered rectangular structure \cite{Shpyrko77} with the proposed Si-Si bond on the surface. Shpyrko and co-workers \cite{Shpyrko77} claimed that the Au-Si surface ordering was induced by surface-induced freezing \cite{PLUIS1990282}, but the ordered prefreezing surface  will  only appear near the transition temperature rather than a  wide range of  temperatures,  inconsistent with  the case of Au-Si. Moreover, density functional theory (DFT) calculations proved that the Si-Si bond is energetically unfavorable in the Au-Si alloy \cite{PhysRevB.81.140202,PhysRevB.95.134109}, implying the inconsistency in literature. This ordered surface plays a key role in  crystal growth and solidification \cite{doi:10.1021/acsnano.1c00500,https://doi.org/10.1002/adma.201806544}, but the atomic arrangement is still unclear. Therefore, uncovering the atomic structures of surface crystallization and Au-Si alloy will be beneficial for understanding exotic properties of Au-Si metallic glass.

In this article, $ab ~ initio$ molecular dynamics (AIMD) simulations  were performed to investigate the bonding features of Au-Si in the undercooled state. The partial distribution function and coordination number analysis were employed to explore the SRO of liquid Au-Si alloy, and a set of  \ce{Au8Si} dodecahedrons with local five-fold symmetry are  the most energetically favorable motifs. The connection modes of different \ce{Au8Si} motifs with similar energies are substantial to reconstruct the Au-Si eutectic alloy, subtly affecting the structural stability of Au-Si. With dimensionality reduction,  the basic building motifs, i.e., \ce{Au8Si} dodecahedrons, are reduced to be  \ce{Au5Si} pyramids, which  are Si-centered pentagons with five-fold symmetry.  This phenomenon is confirmed by a systematic investigation of surface structures on the Au-Si alloy. Moreover,  a  2D ordered   \ce{Au2Si} monolayer composed of \ce{Au5Si} pyramids is found, which is more energetically favorable compared with  all the earlier proposed surface structures.  Notably, the simulated scanning tunneling microscope (STM) images are in much better agreement with the experimental results. Furthermore, the remarkable difference of structural similarity between the Au-Si bulk and surface structures is uncovered, providing insights into understanding  why the  bulk structure is a metallic glass while the surface alloy exhibits long-range order.

\section{Results and discussion}

\begin{figure}[!t]
\centering
\includegraphics[width=\linewidth]{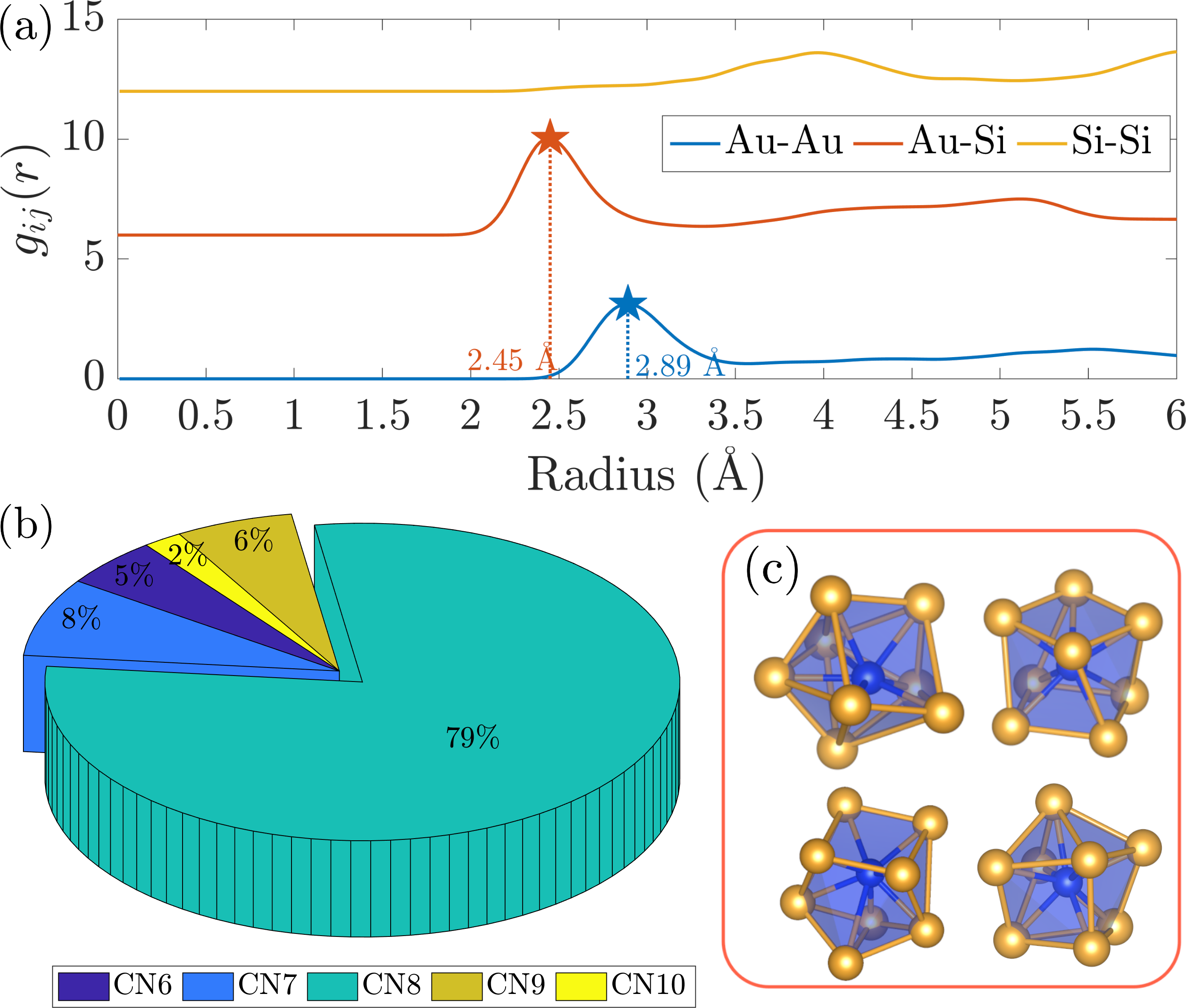}
\caption{(a) The pair-correlation functions of  \ce{Au82Si18} BMG at room temperature. For clarification, the Si-Si and Au-Si partials are shifted by 5 and 10, respectively. (b) The coordination number distribution of  \ce{Au82Si18} BMG at room temperature. (c) The most frequent \ce{Au8Si} motifs in  \ce{Au82Si18} BMG.}
\label{fig1}
\end{figure}

To extract the inherent local SRO of  \ce{Au82Si18}, the initial configurations were constructed  based on  the Au bulk in thermal equilibrium, where a certain number of  Au atoms were randomly replaced  with Si atoms. To avoid the systematic error, 100 configurations were used to produce structural information, for instance, the average coordination numbers and the partial pair-correlation functions (see the details in the section of computational methods in the Supplemental Material (SM) \cite{SM}). As shown in Fig. \ref{fig1}(a), the partial pair-correlation functions $g_{ij}(r)$  of  \ce{Au82Si18} were obtained after the undercooling process, where  the simulation temperatures were quickly reduced to 300 K from 2000 K with a large cooling rate of $1\times10^{11}~ \textrm{K/s}$. The first peak of Au-Si is located at around 2.45 \AA, suggesting the intensive mixing between Au and Si atoms. The first broad peak of Si-Si appears around 4 \AA, showing that Si atoms are prohibited from being bonded together and surrounded by Au atoms. This phenomenon implies the existence of close packing that holds Si atoms like ``solute" atoms in the center of polyhedrons. The coordination number (CN) analysis reveals that \ce{Au8Si} polyhedrons are the most frequent motifs in the Au-Si BMG, where the ratio reaches up to $79 \%$, as shown in Fig. \ref{fig1}(b). It is worth noting that the frequent \ce{Au8Si} motifs have local five-fold symmetry in Fig. \ref{fig1}(c), which is responsible for the SRO. As a result, we can infer that the atomic structure of the Au-Si alloy is the self-assembly of \ce{Au8Si} dodecahedrons by connecting each other in a close-packed manner.


\begin{figure}[!t]
\centering
\includegraphics[width=\linewidth]{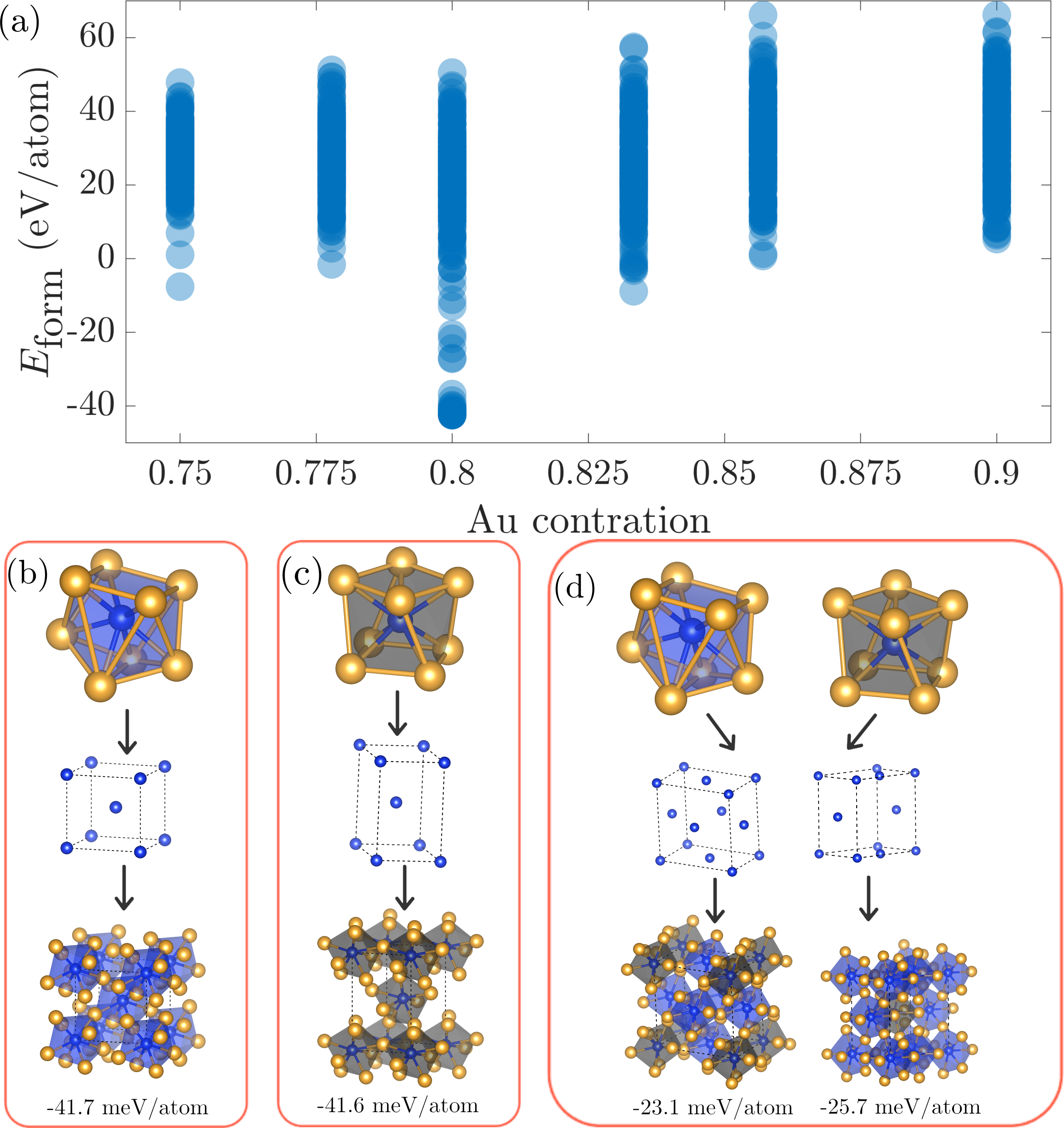}
\caption{(a) The formation energy of Au-Si alloy with various Au concentrations. (b), (c) The most stable \ce{Au4Si} crystal structures, which are constructed by  the fundamental \ce{Au8Si} motifs with different  corner sharing modes. (d) The structure constructed from the two fundamental \ce{Au8Si} motifs with other corner sharing modes. Si and Au atoms are represented by blue and golden spheres, respectively.}
\label{fig2}
\end{figure}

To determine the MRO in  Au-Si, we need to reveal the origin of its structural stability, where the building blocks and connection rules are crucial to identify the local atomic ordering.   According to the bonding features we obtained in Figs. \ref{fig1}(a), \ref{fig1}(b), we constructed a series of Au-Si configurations, which contain  Au$_n$Si ($n=6-10$) polyhedrons in a corner-sharing or edge-sharing mode, to explore stable structures of Au-Si alloy.  As shown in Fig. \ref{fig2}(a), the \ce{Au4Si} structure has the lowest formation energy, which is in excellent agreement with the eutectic point of the Au-Si phase diagram \cite{Okamoto1983}. Particularly, two most stable structures with almost the same energy have been found, which are both comprised of \ce{Au8Si} motifs in body-centered cubic (see Fig. \ref{fig2}(b)) and hexagonal close-packed (see Fig. \ref{fig2}(c)) structures. Moreover, we find that these two \ce{Au8Si} motifs can be combined together to form  relatively stable structures with a larger number of atoms (see Fig. \ref{fig2}(d)), indicating that the connection modes are not dominant for the structural stability of Au-Si compared to the local atomic ordering. Therefore,  the MRO at the atomic level is ascribed to the interconnection modes of \ce{Au8Si} motifs. It is worth noting that the  formation energy distribution at Au concentration of 0.8 is very close, implying a flat  energy landscape  and  the origin of formation mechanism in  Au-Si \cite{PhysRevLett.112.083401}.

\begin{figure}[!t]
\centering
\includegraphics[width=\linewidth]{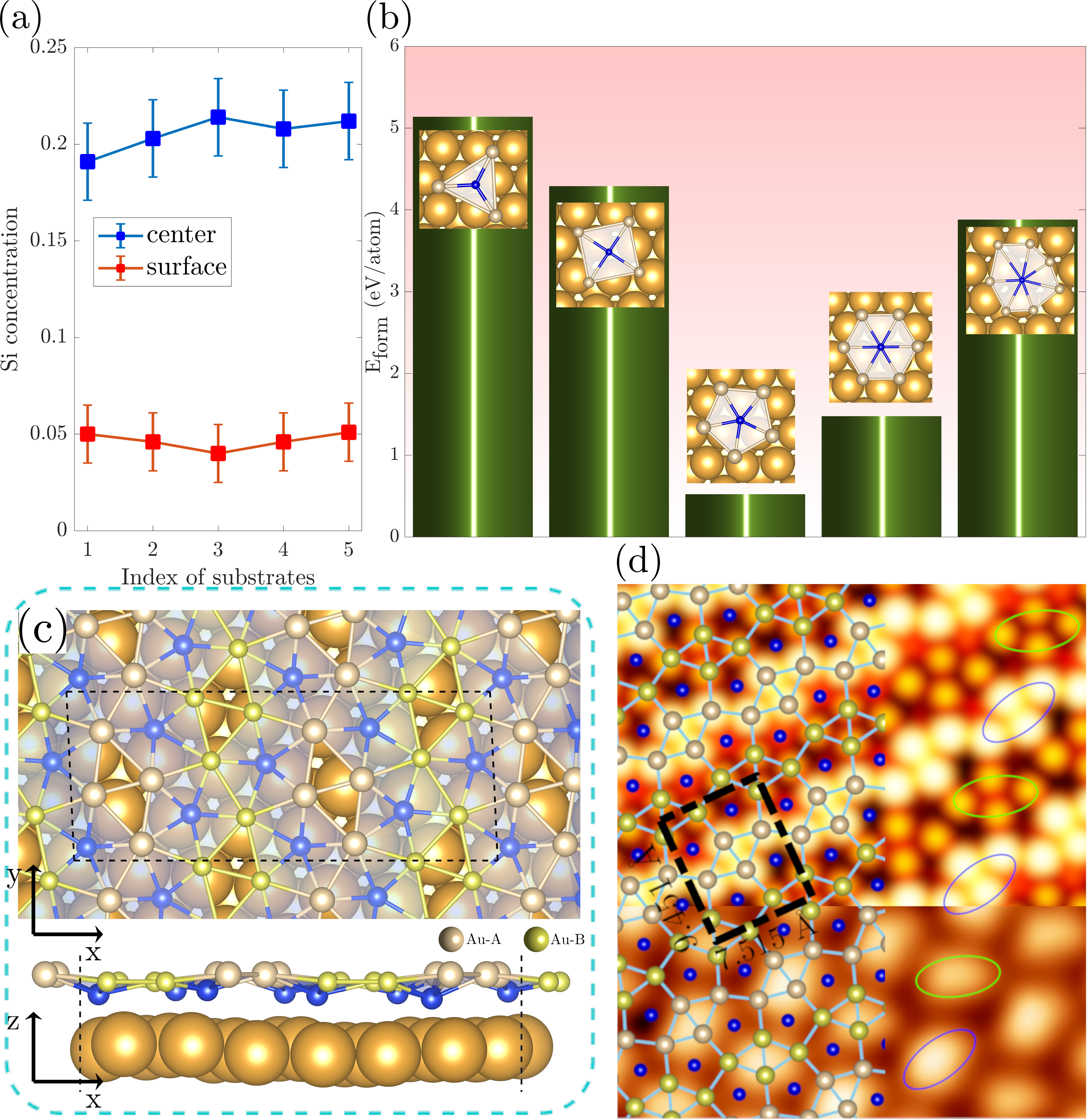}
\caption{(a) The concentration of Si atoms in the surface layer and central region for five different slab models.  (b) Formation energy of one Si atom on the Au(111) substrate. The insets show the corresponding structures. (c) The top view and side view of the stable slab model. Surface Si atoms are represented by blue spheres, and the black dashed lines mark the lattice of the slab. (d) The simulated (upper) and experimental (bottom) STM images. The experimental STM image is reprinted with permission from Ref. \cite{https://doi.org/10.1002/adfm.201906053}.}
\label{fig3}
\end{figure}

The \ce{Au8Si} dodecahedrons with local five-fold symmetry are found to be the SRO in Au-Si,  but with dimensionality reduction the SRO of surface structures still leaves unknown.  The surface crystallization of  Au-Si BMG was observed in the last decade, however, the origin and atomic structure of the surface ordering are still in debate \cite{Shpyrko77,https://doi.org/10.1002/adma.201806544,PLUIS1990282}. We carried out AIMD simulations at 650 K  to study the Si content in a roughly 25 \AA-thick \ce{Au82Si18} slab with six layers, and  five different slab models were considered to reduce systematic errors. During AIMD simulations Si atoms migrate to the surface from the central region, resulting in the higher concentration of Si atoms (around $42\%$) in the surface region, as shown in Fig. \ref{fig3}(a). Therefore, our results agree with the fact that Si atoms prefer the surface segregation in experiments \cite{Shpyrko77}, and the surface structure of Au-Si alloy is the mixture of Au and Si atoms.

Interestingly, this ordered rectangular surface structure with the same lattice parameters has also been experimentally observed by depositing Si atoms on the  Au(111) substrate \cite{Sadeddine2017,https://doi.org/10.1002/adfm.201906053}. Therefore, it is possible to search for the stable Au-Si surface structure on the Au(111) substrate. At the initial stage of the nucleation process, only one Si atom was introduced to react with the Au(111) substrate. The Au$_n$Si ($n=3-7$) clusters with various coordination numbers for Si on Au(111) are shown in Fig. \ref{fig3}(b).  It is apparent that the \ce{Au5Si} cluster with five-fold symmetry has the lowest formation energy, indicating that the \ce{Au5Si} pyramid is an energetically favorable motif in the Au-Si system.

To efficiently search for the possible structure of 2D Au-Si alloy, surface structure generation  methods were employed to fertilize the structure database of Au-Si monolayers. We generated non-duplicated 2D Au-Si surface structures with various atomic ratios and distributions by the package of Structures of Alloy Generation And Recognition (SAGAR)  \cite{doi:10.1021/acs.jpca.0c02431,HE110386} and the crystal prediction algorithm RG2 based on space group theory and graph theory  \cite{PhysRevB.97.014104}.  To consider the lattice mismatch between monolayers and Au(111), we constructed  various supercells.  High-throughput first-principles calculations were performed to  determine the possible surface structure on Au(111). As shown in Fig. S1 of the SM \cite{SM}, our results point out that Au-Si monolayers with \ce{Au5Si} pyramids are rather stable compared with those monolayers without \ce{Au5Si} pyramids, indicating that the \ce{Au5Si} motif with five-fold symmetry is the key SRO in surface structures of Au-Si.

Particularly, the most stable Au-Si monolayer on  Au(111) is \ce{Au2Si} comprised of pentagons and rhombuses, and the relaxed \ce{Au2Si} monolayer on Au(111) are shown in Fig. \ref{fig3}(c). Intriguingly, the lattice parameters of the most stable \ce{Au2Si} monolayer are $a = 9.451$ {\AA} and $b = 7.515$ \AA, which are in excellent agreement with previous experiment results  \cite{Shpyrko77,Sadeddine2017,https://doi.org/10.1002/adfm.201906053}.
Si atoms are at a lower height than Au atoms in the Au-Si monolayer, implying that each rhombus contains four Au atoms in the topmost layer and Si atoms are separated by these \ce{Au4} rhombuses. According to the experimental and simulated STM images shown in Fig. \ref{fig3}(d), the bright protrusions correspond to the \ce{Au4} rhombuses.  There are two types of protrusions in the \ce{Au2Si} monolayer that are rotated with respect to each other as highlighted by the ellipses  \cite{Sadeddine2017}, corresponding to the two types of Au atoms with different heights in Fig. \ref{fig3}(c).  The \ce{Au5Si} pyramids reflect dark pores in the STM image because the Si atoms are at a relatively low height. Therefore, we conclude that the \ce{Au2Si} monolayer forms by depositing Si atoms on the Au(111) substrate.

To further confirm the existence of the ordered \ce{Au2Si} monolayer at the surface of \ce{Au82Si18} BMG, various Au-Si overlayers were deposited on the five slab models mentioned above in Fig. \ref{fig3}(a). The calculated formation energies on these five substrates are shown in Fig. S2 of the SM \cite{SM}, where the \ce{Au2Si} overlayer exhibits robust stability on different substrates, while other overlayers are less stable. This is strong evidence that the \ce{Au2Si} monolayer is the pretty stable ordered surface structure of Au-Si BMG.

The Au-Si BMG displays diverse structural characteristics in different dimensions, which is attributed to  unusual Au-Si bonding properties and dimensionality reduction  \cite{Shpyrko77}.  To unveil the deep physical nature of the phenomenon, the energy landscape of the Au-Si system is  comprehensively investigated to  uncover the origin of the distinction between bulk and surface. If the global minimum is surrounded by  numerous local  minima, namely, lots of low energy  saddle points are  around the ground state structure, this system is a glassy system, otherwise, it belongs to a structure seeker system  \cite{PhysRevLett.112.083401}.  For instance, the carbon and boron nitride systems can be clearly identified as structure seekers while the large boron cluster systems are glassy  \cite{PhysRevLett.106.225502}.

\begin{figure}[!t]
\centering
\includegraphics[width=\linewidth]{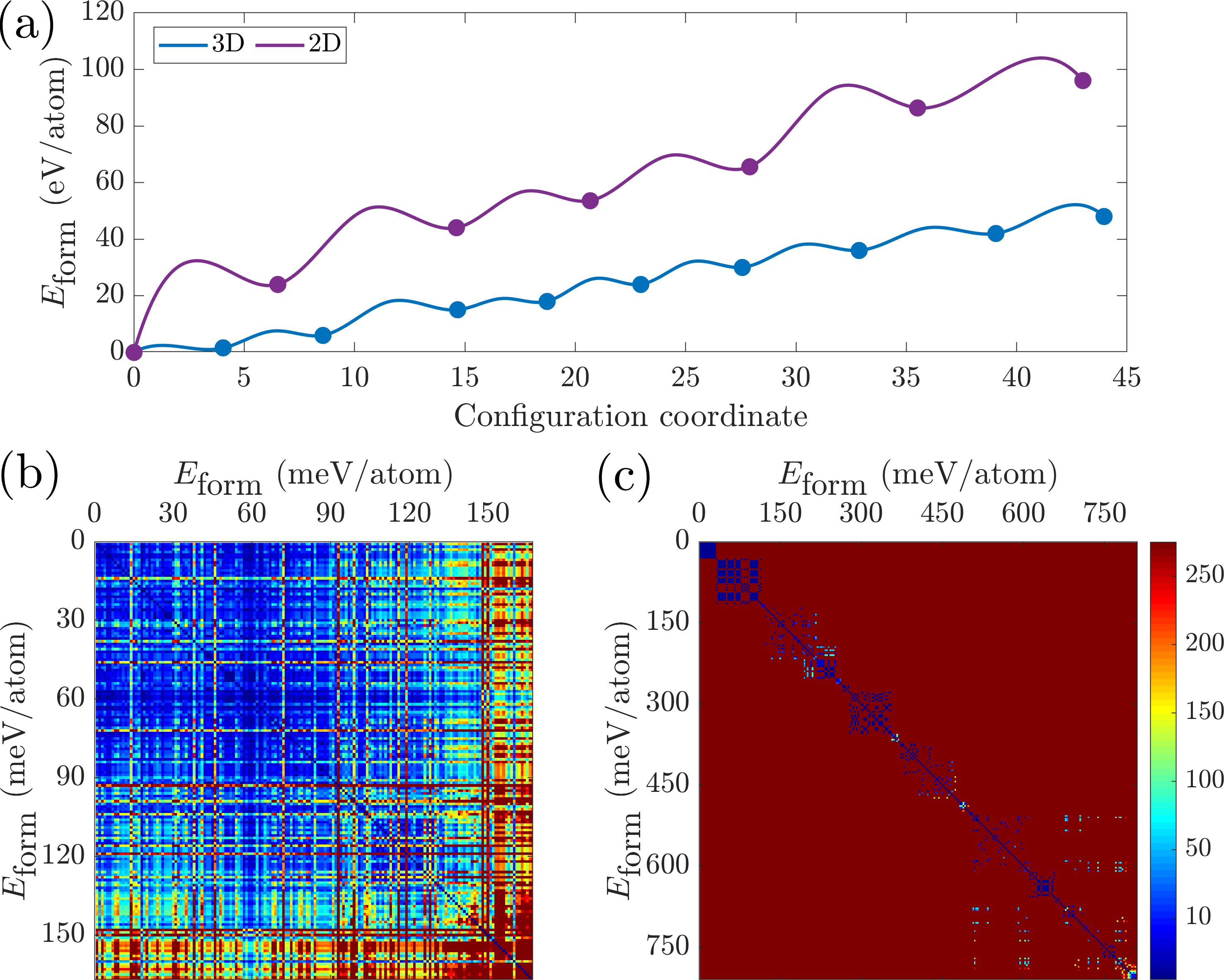}
\caption{(a) The distance-energy analysis for low-energy  Au-Si structures in 2D and 3D systems. The structural similarity distribution in bulk and surface Au-Si are shown in (b) and (c), respectively. The colors represent the similarity degree between two structures, where blue/red represents the small/large distance, corresponding to the  shared colorbar in the right, which is applicable for both 3D and 2D systems.}
\label{fig4}
\end{figure}

To define the distance between any two structures, we employed the Smooth Overlap of Atomic Positions (SOAP) descriptor  \cite{PhysRevB.87.184115}  to encode regions of atomic geometries by using local expansion of a gaussian smeared atomic density.  The bulk Au-Si alloy and surface structures were generated by SAGAR and RG2 algorithm as mentioned above. We chose several low-energy  Au-Si structures to perform the distance-energy analysis  \cite{doi:10.1063/1.4828704}  in 2D and three-dimensional (3D) systems as shown in Fig. \ref{fig4}(a).   For 2D surface structures,  the  formation energy increases rather quickly  with configuration coordinate,  and the saddle  points are relatively far away from each other according to the configuration coordinate compared to the case of 3D bulk structures.   In the bulk case, the blue color in Fig. \ref{fig4}(b) represents that the distances among the low energy structures are rather small, indicating that there is a low energy barrier between two local minima. This feature leads to the fact that it is difficult to form any ordered bulk Au-Si alloy during the rapid cooling process, which is confirmed by  molecular dynamics simulations in Au-Si BMG  \cite{RAN2021120787}. However, the image of the Au-Si surface structure (see Fig.  \ref{fig4}(c)) is almost filled with red color, implying that distances are much larger between low and high energy structures.  Therefore, the long-range ordering can easily appear in surface structures.

In summary, we have uncovered  the local ordering of Au-Si, where the  \ce{Au8Si} motifs with local five-fold symmetry are the basic building blocks and the different connection modes  reveal the MRO. With dimensionality reduction, the SRO of Au-Si surface structure has been transformed into \ce{Au5Si} pyramids with five-fold symmetry. We have concluded that the \ce{Au2Si} monolayer composed of \ce{Au5Si} pyramids is the  surface structure observed in experiments, due to the lowest formation energy  and excellent agreement with STM images. Employing the structural similarity matrix based on the SOAP descriptor  to metricize the distance among Au-Si structures, we have shown that there is a smooth energy landscape with dense local minima for the Au-Si bulk, while the energy landscape of Au-Si surface structures is relatively steep around the ground state.  Our work not only supplies a structure search method to determine the  Au-Si structures in different dimensions but also expounds on the intrinsic origin of the order-disorder transition in Au-Si alloy from 2D to 3D, which may pave a promising avenue to explore the novel properties in other metallic glasses.




\begin{acknowledgement}
This work was supported by the Science, Technology, and Innovation Commission of Shenzhen Municipality (Grant Nos. RCYX20200714114523069 and ZDSYS20190902092905285), the National Natural Science Foundation of China (No. 11974160), and Key-Area Research and Development Program of Guangdong Province (No. 2020B010183001). The computer time at the Center for Computational Science and Engineering at Southern University of Science and Technology is gratefully acknowledged.

\end{acknowledgement}




\bibliography{acs-achemso}

\end{document}